# Social Media and COVID-19: Can Social Distancing be Quantified without Measuring Human Movements?

Mackenzie Anderson, Amir Karami, Parisa Bozorgi

**Abstract**

The COVID-19 outbreak has posed significant threats to international health and the economy. In the absence of treatment for this virus, public health officials asked the public to practice social distancing to reduce the number of physical contacts. However, quantifying social distancing is a challenging task and current methods are based on human movements. We propose a time and cost-effective approach to measure how people practice social distancing. This study proposes a new method based on utilizing the frequency of hashtags supporting and encouraging social distancing for measuring social distancing. We have identified 18 related hashtags and tracked their trends between Jan and May 2020. Our evaluation results show that there is a strong correlation (P<0.05) between our findings and the Google social distancing report.

**Introduction**

The World Health Organization (WHO) announced the outbreak of COVID-19 to be a public Health Emergency in January 2020 (WHO, n.d.). As of May 30, 2020, the number of positive cases was more than 6 million globally, with over 369,000 deaths[1]. The outbreak has posed significant threats to international health and the economy. In the absence of treatment for this virus, public health officials asked the public to practice social distancing to reduce the amount of physical contact. In addition, social distancing helps to flatten the curve of positive COVID-19 cases. However, people have a desire for social interactions (Cohen, 2004).

While all US states instituted social distancing measures, measuring social distancing is a challenging task. In addition, developing traditional methods (e.g., survey) is costly and time-consuming. Some companies and institutes such as Google have proposed methods to quantify social distancing based on measuring distances and movements.

Social media has become a mainstream channel of communication where users share and exchange information(Sha et al., 2020). In 2019, 72% of U.S. adults use at least one social media site[2]. In the last decade, social media platforms have grown in popularity, and now Facebook, Twitter, and Instagram are readily available on mobile devices, continuously connecting users to a stream of information. In public health surveillance, social media can provide communication in real time and at relatively low cost, monitor public response to health issues, track disease outbreak and infectious disease. Public health experts have investigated social media for different health issues such as diet, diabetes, and obesity (Karami et al., 2018).

Social distancing was measured by calculating movements shared on social media during the COVID-19 epidemic (Xu et al., 2020). While the current methods are based on measuring human movements, this study proposes a new method based on utilizing the frequency of tweets supporting and encouraging social distancing for measuring social distancing.

---

[1] https://www.worldometers.info/coronavirus/
[2] https://www.pewresearch.org/internet/fact-sheet/social-media/

**Methodology**

This research proposed a simple method without measuring human movements. The method is based on the idea of measuring the frequency of tweets supporting and encouraging social distancing. The hashtags were identified based on qualitatively searching hashtags in Twitter, Google, and websites (e.g., https://www.digitaltrends.com/social-media/all-the-hashtags-you-need-to-know-about-social-distancing/). We selected the hashtags that had a higher chance to be supportive and encouraging tweets. This process provided a list of 18 hashtags such as #StayHome utilized to support or encourage social distancing (Table 1). For example, an American actor, Morgan Freeman, posted a tweet containing #StayHome. We have collected the frequency of the hashtags in US from Jan, 2020 to May 26, 2020. We used Brandwatch, a third-party social media data provider, to obtain the data. The data will be available on the web site of the first author.

| | | |
|---|---|---|
| #StaySafeStayHome | #StayHomeSweetHome | #Lockdown2020 |
| #SocialDistancing | #StayHomeSaveLives | #Quarantine |
| #SocialDistancingWorks | #HealthyAtHome | #Quarantined |
| #FlattenTheCurve | #Lockdown | #Quarantine2020 |
| #StayHome | #LetsStayHome | #Quaranthriving |
| #StayAtHome | #TogetherAtHome | #Quarantining |

Table 1: Twitter Hashtags for Supporting or encouraging Social Distancing

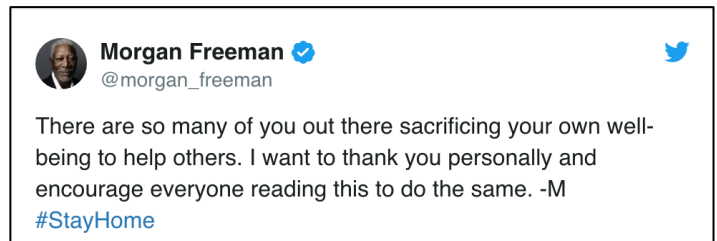

Figure 1: A Tweet Encouraging Social Distancing

**Evaluation**

To evaluate our method, we utilized a correlation analysis using Pearson method to find the correlation between our approach and the COVID-19 Community Mobility Report released by Google. We selected the Google measure because millions of people use Google's applications such as Google's map. The Google Community Mobility Report[3] provides insights into what has changed in response to policies aimed at combating COVID-19. This report shows movement trends over time by geography, across different categories of places such as retail and recreation, groceries and pharmacies, parks, transit stations, workplaces, and residential. We used the correlation function in R to investigate the correlation between the number of hashtags and the recorded movement by Google in different places.

**Results**

Our findings show that the number of hashtags had an increasing trend in March and then a decreasing trend in April and May. Figure 3 shows the total frequency of hashtags with the highest and lowest frequency for #StayHome and #Quaranthriving, respectively.

---

[3] https://www.google.com/covid19/mobility/

Our evaluation analysis shows a meaningful correlation ($P < 0.05$) between the total frequency of the 18 hashtags and the Google movement trends (Figure 4). This evaluation shows that there is a positive correlation ($R > 0$) between the frequency of hashtags and residential movements and a negative correlation ($R < 0$) between the frequency of tweets and non-residential movements. Our findings show that US people used more supportive and encouraging hashtags for social distancing when they were at residential places and practiced social distancing and utilized less supportive and encouraging hashtags for social distancing when they were not at residential places. So, the frequency of hashtags can be used for quantifying social distancing during the COVID-19 pandemic.

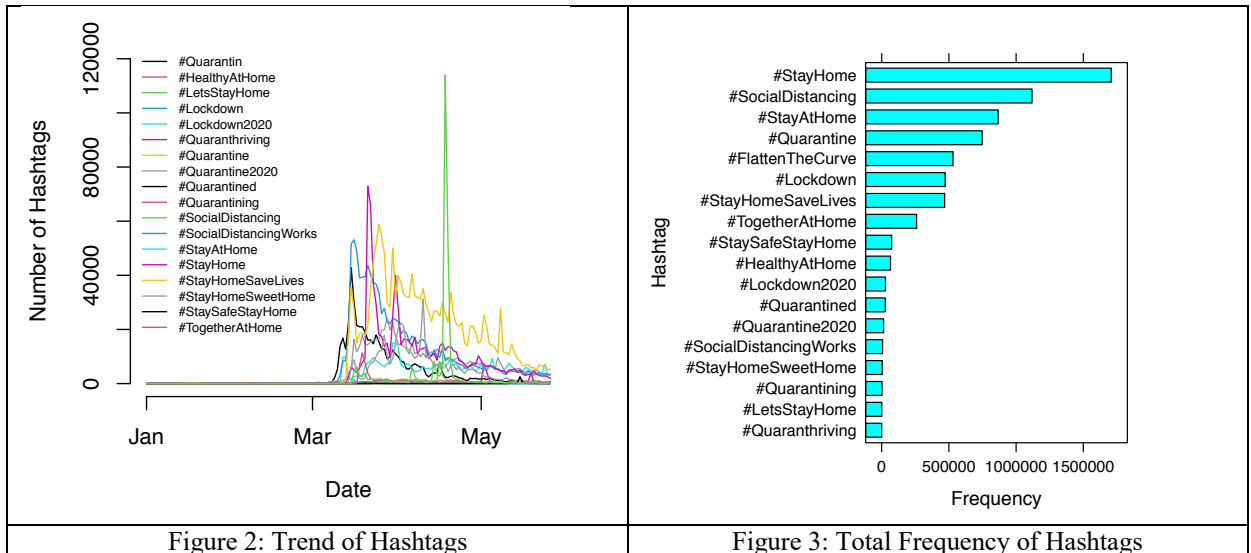

| Figure 2: Trend of Hashtags | Figure 3: Total Frequency of Hashtags |
| --- | --- |

**Discussion and Conclusion**

This study proposed a cost and time effective approach to measure social distancing during the COVID-19 pandemic. Our findings show a strong correlation between the frequency of hashtags supporting and encouraging social distancing and the Google social distancing report. This research offers a new dimension for social distancing analysis during a pandemic and assessing social distancing policies.

While this paper provided a new insight on measuring social distancing, it has some limitations. First, we didn't collect non-English tweets posted in US. Second, the evaluation process was limited to a national level analysis. Third, we selected one source for comparison. Future work could utilize non-English tweets, extend the evaluation process to state and county levels, and use other social distancing calculations.

**Acknowledgement**

This work is partially supported by the College of Information and Communication Internal Collaboration Grant and the Big Data Health Science Center (BDHSC) at the University of South Carolina. All opinions, findings, conclusions and recommendations in this paper are those of the authors and do not necessarily reflect the views of the funding agency.

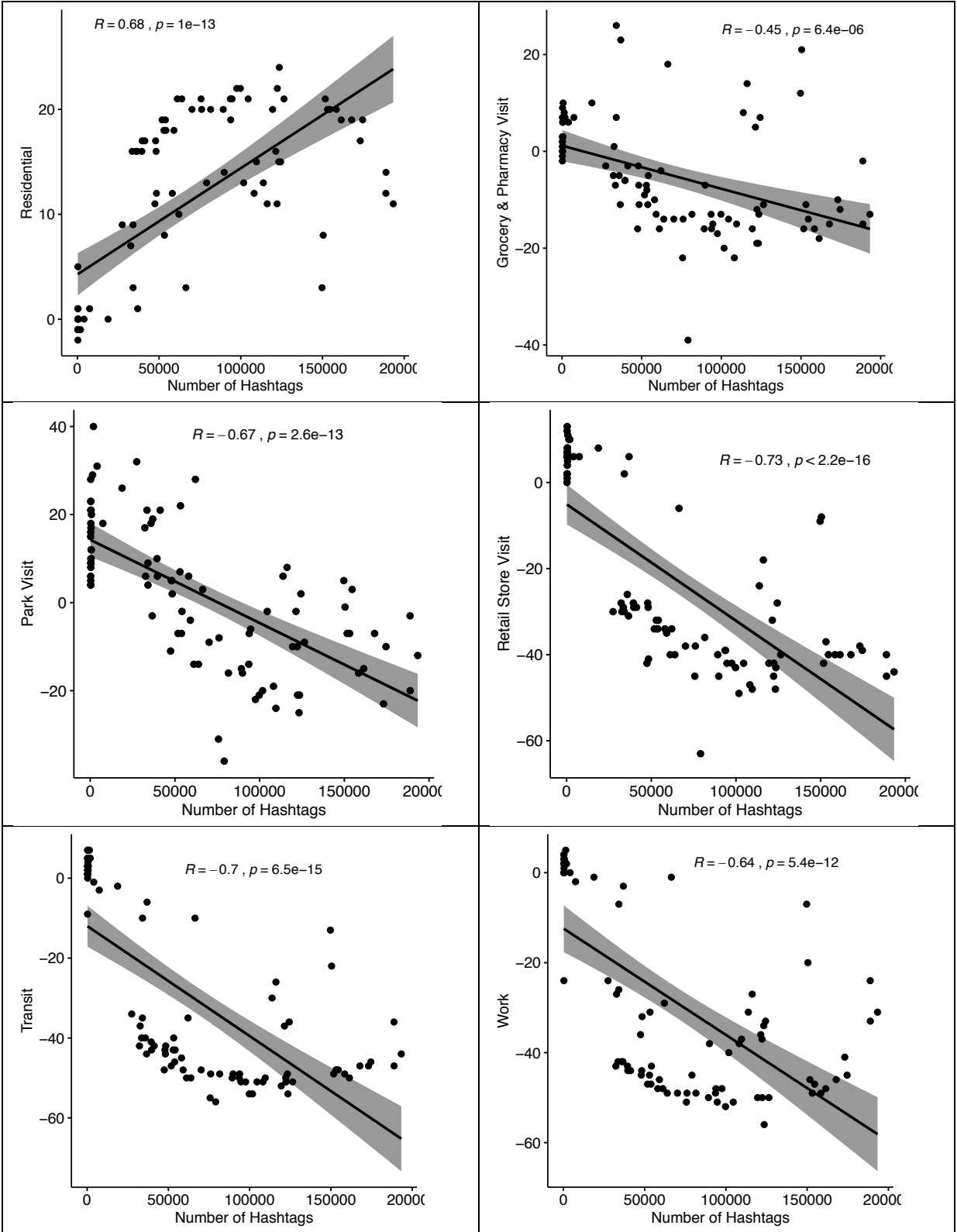

Figure 4: Pearson Correlation Analysis: Number of Hashtags vs Google COVID-19 Community Mobility Report

**References**

Cohen, S. (2004). Social relationships and health. *American Psychologist*, *59*(8), 676.

Karami, A., Dahl, A. A., Turner-McGrievy, G., Kharrazi, H., & Shaw Jr, G. (2018). Characterizing diabetes, diet, exercise, and obesity comments on Twitter. *International Journal of Information Management*, *38*(1), 1–6.

Sha, H., Hasan, M. A., Mohler, G., & Brantingham, P. J. (2020). Dynamic topic modeling of the COVID-19 Twitter narrative among US governors and cabinet executives. *ArXiv Preprint ArXiv:2004.11692*.

WHO. (n.d.). *Rolling updates on coronavirus disease (COVID-19)*. https://www.who.int/emergencies/diseases/novel-coronavirus-2019/events-as-they-happen#:~:text=The%20outbreak%20was%20declared%20a,on%2030%20January%202020.

Xu, P., Dredze, M., & Broniatowski, D. A. (2020). The twitter social mobility index: Measuring social distancing practices from geolocated tweets. *ArXiv Preprint ArXiv:2004.02397*.